\expandafter\def\csname ver@fixltx2e.sty\endcsname{} 
\PassOptionsToPackage{pdfpagelabels=false}{hyperref} 
\documentclass[fleqn,usenatbib,useAMS]{mnras}

\usepackage[T1]{fontenc}
\usepackage{graphicx}
\usepackage{ae,aecompl}
\usepackage{newtxtext,newtxmath,units}
\usepackage{xspace}

\usepackage{siunitx}
\DeclareSIUnit \parsec {pc}
\DeclareSIUnit\year{yr}

\usepackage[nolist]{acronym}
\newacro{SNR}[SNR]{signal to noise ratio}
\newacro{PSD}[PSD]{power spectral density}

\newcommand{\cqg}{Class. Quantum Gravity}
\newcommand{\prx}{Phys. Rev. X}

\newcommand{\pastro}{\ensuremath{p_{\rm astro}}}
\newcommand{\odds}{\mathcal{O}}
\newcommand{\likelihood}{\mathcal{L}}
\newcommand{\signal}{\mathcal{S}}
\newcommand{\noise}{\mathcal{N}}
\newcommand{\xiGH}{\xi_{g}^{\textsc{h}}}
\newcommand{\xiGHhat}{\hat{\xi}_{g}^{\textsc{h}}}
\newcommand{\xiGL}{\xi_{g}^{\textsc{l}}}
\newcommand{\xiGLhat}{\hat{\xi}_{g}^{\textsc{l}}}
\newcommand{\Msun}{\textrm{M}_{\sun}}
\newcommand{\Omicron}{\textsc{Omicron}\xspace}
\newcommand{\result}{\ensuremath{0.03}}

\title[The astrophysical odds of GW151216]{The astrophysical odds of GW151216}
\author[G. Ashton \& E. Thrane]{
Gregory Ashton$^{1,2,}$\thanks{E-mail: gregory.ashton@ligo.org},
Eric Thrane$^{1,2,}$
\\
$^{1}$School of Physics and Astronomy, Monash University, Vic 3800, Australia,\\
$^{2}$OzGrav: The ARC Centre of Excellence for Gravitational Wave Discovery, Clayton VIC 3800, Australia
}

\begin{document}

\label{firstpage}
\pagerange{\pageref{firstpage}--\pageref{lastpage}}
\maketitle

\begin{abstract}
The gravitational-wave candidate GW151216 is a proposed binary black hole event from the first observing run of the Advanced LIGO detectors. Not identified as a bona fide signal by the LIGO--Virgo collaboration, there is disagreement as to its authenticity, which is quantified by $p_\text{astro}$, the probability that the event is astrophysical in origin. Previous estimates of $p_\text{astro}$ from different groups range from 0.18 to 0.71, making it unclear whether this event should be included in population analyses, which typically require $p_\text{astro}>0.5$. Whether GW151216 is an astrophysical signal or not has implications for the population properties of stellar-mass black holes and hence the evolution of massive stars. Using the astrophysical odds, a Bayesian method which uses the signal coherence between detectors and a parameterised model of non-astrophysical detector noise, we find that $p_\text{astro}=\result{}$, suggesting that GW151216 is unlikely to be a genuine signal. We also analyse GW150914 (the first gravitational-wave detection) and GW151012 (initially considered to be an ambiguous detection) and find $p_\text{astro}$ values of 1 and 0.997 respectively. We argue that the astrophysical odds presented here improve upon traditional methods for distinguishing signals from noise.
\end{abstract}


\setlength{\abovedisplayskip}{3pt}
\setlength{\belowdisplayskip}{3pt}

\section{Introduction}
\label{sec:introduction}
Transient gravitational wave-astronomy has opened a new window with which to study black holes and neutron stars.
The LIGO \citep{LIGO} and Virgo \citep{virgo} collaborations have now completed three observing runs and announced 13 binary coalescence signals \citep{GWTC1, GW190425, GW190412}.
The data collected by these observatories is public allowing independent groups to reaffirm observations and identify new candidates \citep{Zackay2019, Nitz2019, Venumadhav2018, Venumadhav2019}.

In addition to astrophysical signals, gravitational-wave detector data contains transient non-Gaussian noise artefacts, often referred to as glitches \citep{2008CQGra..25r4004B, transient_noise, nutall2015, cabero2019, powell2018}. Glitches degrade our ability to identify signals, i.e. the sensitivity of the detector; when the cause of the glitch is fully understood, the optimal solution is to remove the data containing the glitches which improves the sensitivity of the detector \citep{2018CQGra..35f5010A}. However, the cause of many glitches is not understood and hence they cannot be removed from the data, but must be treated as part of the background noise of the detector.
Traditional search methods (see, e.g. \citet{cannon2013, usman2016} and \citet{capano2017} for a review of the methods) deal with this by estimating the background using bootstrap methods~\citep{Efron}.
Bootstrap methods are defined by the use of an empirical distribution to estimate a quantity of interest.
Subsequently, candidates are assigned an astrophysical probability, \pastro, based on the empirical output of the search pipeline; see \citet{2016ApJ...833L...1A, 2016ApJS..227...14A, GWTC1} for details.
For loud events such as the GW150914 \citet{GW150914}, the first observed binary black hole coalescence, $\pastro \approx 1$.
Meanwhile, for marginal candidates, $\pastro \in [0.5, 0.99]$. 
Different search pipelines produce different values of $\pastro$ due to differing assumptions.
For loud events, this is of little consequence, but as we will see, understanding these assumptions can be crucial for marginal candidates.

GW151216 was reported as a significant trigger in O1, the first LIGO observing run, with $\pastro=0.71$ by \citet{Zackay2019}. The event
was not included in the first LIGO--Virgo gravitational-wave transient catalogue covering the O1 and O2 observing runs \citep{GWTC1}.
The candidate was also identified in \citet{nitz2020}, but with $\pastro=0.18$, less than the 0.5 threshold used to determine inclusion in the catalogue.
In the original analysis, \citeauthor{Zackay2019} noted the large effective spin of the candidate, which led to a range of implications; e.g.~\cite{piran2020, fragione2020, luca2020}).
However, in a systematic study \citep{Huang2020}, it was shown that support for the effective spin was sensitive to the choice of prior.
We summarise the various significance estimates for all events analysed in this work in Table~\ref{tab:overview}.

In this work, we study the significance of GW151216 using the astrophysical odds \citep{bcr2}.
This method is different from traditional methods in that it eschews bootstrap noise estimation. 
Instead, it directly models and fits for the population properties of glitches as they appear projected onto the parameter space of compact binary coalescence signals.
By combining the notion of glitches as incoherent signals \citep{veitch2010, isi2018}, and using contextual data to measure the population properties of glitches, the astrophysical odds can elevate the significance of marginal candidates based on their coherence between detectors and their properties in the context of typical glitches.
The odds is a Bayesian ratio of probabilities comparing a signal and noise hypothesis complete with prior probability; it can be used to directly weight posteriors in the context of a population analysis, disposing of the need for arbitrary thresholds for inclusion \citep{Galaudage2019,gaebel2019} and can be employed directly to the analysis of multi-messenger events using the framework laid out in \citet{ashton2018}.

In order to give the results for GW151216 context, and to validate our method, we also analyse two other binary black hole signals: GW150914, the first and most significant signal in O1 and GW151012, first reported as a ``trigger'' \citep{2016PhRvD..93l2003A} and subsequently upgraded in significance to a candidate \citep{GWTC1, 1OGC}.
In the future, we expect more candidates to be identified in the open data by independent pipelines (see, e.g. \citet{Venumadhav2019}).
While we focus here on GW151216, our broader goal is to establish a unified catalogue, sourced from multiple groups, each event with a single, reliable value of $p_\text{astro}$.
The $p_\text{astro}$ in this unified catalogue will not depend on the search pipeline used to first identify each trigger.

\section{Method}
\label{sec:method}
Following \citet{bcr2}, we use a Bayesian framework to calculate the astrophysical odds, $\odds$.
The odds answers the question: what is the ratio of probability that a $\Delta t= \unit[0.2]{s}$-duration data segment $d_i$ spans the coalescence time\footnote{The coalescence time is defined differently for different waveforms, but it is approximately synonymous with time of peak gravitational-wave amplitude. While a gravitational waveform can span several segments, the coalescence time always falls in just one segment.} of an astrophysical signal versus the probability that it contains noise?
The noise can be either Gaussian or it can include glitch.
The odds for a signal in data segment $d_i$ in a larger data set $d$ are
\begin{equation}
    \odds_{\signal_i/\noise_i}(d) \approx \frac{
    \langle\xi\rangle \likelihood(d_i| \signal_i)}{
    \int  \likelihood(d_i | \noise_i, \Lambda_\noise) \pi(\Lambda_\noise | d_{i \neq k}, I) \, d\Lambda_{\noise}}  \,.
    \label{eqn:bcr2}
\end{equation}
Here, $\xi$ is the probability of a signal in $d_i$ and its expectation value $\langle\xi\rangle$, marginalised over the uncertainty in the astrophysical distribution, is the prior odds; we discuss this more below. 
The term $\likelihood(d_i|\signal_i)$ is the Bayesian evidence (marginal likelihood) for $d_i$ given the signal hypothesis. 
This is the likelihood function commonly used to estimate the parameters of merging binaries, \citep{veitch15}.
Meanwhile, $\likelihood(d_i|\noise_i, \Lambda_\noise)$ is the likelihood of the data given the noise hypothesis. The noise hypothesis is that the data contain either Gaussian noise or non-Gaussian glitches, modelled by uncorrelated (between detectors) binary mergers \citep{veitch2010, isi2018, bcr2}. The noise likelihood is marginalized over $\Lambda_\noise$, a set of hyper-parameters that describe the distributions of glitch parameters.

Finally, $\pi(\Lambda_\noise | d_{i \neq k}, I)$ is the noise parameter prior informed by conditional data $d_{i \neq k}$ and any other cogent information $I$; the importance of this will be made clear later on.
We refer to this distribution as the glitch population properties.
Our present purpose is to describe how we calculate $\odds$ for the three events considered in this paper, so we take Eq.~\ref{eqn:bcr2} as given and refer readers to~\cite{bcr2} for more information including a derivation of $\odds$ and a discussion of the motivation for our noise model.

Equation~\ref{eqn:bcr2} differs slightly from the expression in~\cite{bcr2} because of two simplifying assumptions. First, we assume that the prior signal probability $\xi$ is independent of the glitch hyper-parameters $\Lambda_\noise$. Second we assume that the prior signal probability $\xi \ll 1$ (as expected for astrophysical signals). This allows us to factorise the prior-odds $\pi_{\signal_i/\noise_i}(d_{i\neq k}, I)$ and approximate them by
\begin{align}
\pi_{\signal_i/\noise_i}(d_{i\neq k}, I) = & \frac{\int   \xi \pi(\xi | d_{i\neq k}, I)\, d\xi}{\int (1 - \xi) \pi(\xi| d_{i\neq k}, I)\, d\xi}
\approx \langle \xi \rangle \,,
\label{eqn:prior_odds}
\end{align}
where $\langle \xi \rangle$ is the expectation of value of $\xi$. We omitted $I$ in \citet{bcr2} as all inferences were made from the contextual data alone. In this work, we will make good use of cogent prior information and hence re-introduce it in order to show where this information is important.
With this formalism out of the way, we turn our attention to the evaluation of $\odds$ using data from O1.

The first step is to define the contextual data.
The contextual data is drawn from a span of time near to the candidate of interest.
Ideally, one would like to include as much contextual data as possible, though, not so much that the detector performance is likely to have changed.
A comprehensive study of transient noise in O1 was performed by \citet{transient_noise}.
Using the single-detector burst identification algorithm \Omicron \citep{transient_noise, omicron2}, the rate of all glitches with \ac{SNR}~$>5$ during O1 was found to vary by epoch, but typically was less than $\SI{0.5}{\per\second}$.
However, louder glitches with  \ac{SNR}~$>10$ have a typical rate (excluding vetoed epochs) between $0.01$ and $\SI{0.001}{\per\second}$.
Given this rate, a coincident-observing \SI{24}{\hour} period will contain several thousand quiet glitches and a few hundred loud glitches: a sufficient number to estimate typical population properties.
Thus, we use $\unit[24]{h}$ of contextual data, which is long enough to provide adequate estimates of the population properties of glitches, but short enough to control computational costs.
We define $d_{i \neq k}$ to be the set of \Omicron triggers in the contextual data. 

The next step is to calculate the expectation value of $\xi$. This is often referred to as the ``duty cycle''~\citep{smith18}. 
It is the expectation value for the fraction of segments containing the coalescence time of a gravitational-wave signal.
The duty cycle is straightforwardly related to the local merger rate $R$ and the average time between mergers in the Universe $\tau$: $\xi \sim R \sim \tau^{-1}$.

By assuming a plausible cosmological model,\footnote{These error bars don't include systematic uncertainty associated with the cosmological model, which might increase the uncertainty by a factor of~$\sim2$.} \citet{GW170817_stochastic} obtained $\tau=223^{+352}_{-115} \si{\second}$ based on a local merger rate of $R=103.2^{+110}_{-115}$\si{\per\giga\parsec\cubed\per\year}.
Since then, \citet{O1O2RAP} updated the estimated local merger rate to be $R=53.2^{+58.5}_{-28.8}$\si{\per\giga\parsec\cubed\per\year}.
Combining these results, we obtain a point estimate of
\begin{equation}
    \widehat\xi \approx 4.5\times10^{-4} 
    \left(\frac{\Delta T}{\unit[0.2]{s}}\right)
    \left(\frac{R}{\unit[59]{Gpc^{-3}yr^{-1}}}\right) .
\end{equation}
We approximate the posterior for merger rate as a log-normal distributions, centred on $\widehat\xi$, with shape parameters estimated by fitting the 90\% credible intervals given above.
Using these fits, we Monte-Carlo sample the distribution $\pi(\xi | I)$ in Eq.~\ref{eqn:prior_odds} (we are dropping the contextual data $d_{i\neq k}$ in deference to the information $I$ used above) and find that $\langle \xi \rangle=7.4\times10^{-4}$.
Thus, roughly one in $1/\langle\xi\rangle\approx1400$ segments contains a coalescence time.
For the odds to favour a signal hypothesis, the astrophysical Bayes factor (i.e. all the terms in Eq.~\eqref{eqn:bcr2} \emph{except} $\langle \xi \rangle$) must be larger than these prior odds.

The next step is to estimate the glitch population properties $\pi(\Lambda_\noise | d_{i\neq k}, I)$. We write the set of glitch hyper-parameters as $\Lambda_\noise\equiv\{\xiGH, \xiGL, \lambda_{\noise}\}$ where $\xiGH$ and  $\xiGL$ are the prior probability for a glitch in the LIGO Hanford and Livingston detectors and $\lambda_{\noise}$ is the remaining set of hyper-parameters describing the glitch population properties. Making the simplifying assumption that these are independent, we can write $\pi(\Lambda_\noise | d_{i\neq k}, I)=\pi(\xiGH | d_{i\neq k}, I) \pi(\xiGL | d_{i\neq k}, I)\pi(\lambda_\noise | d_{i\neq k}, I)$.

A computationally efficient means to infer $\lambda_\noise$ is to use the  \ac{SNR}>5 \Omicron triggers present in the $\unit[24]{h}$ span of contextual data as a representative sample of glitches (we pre-filter this list to only include triggers with frequencies between 20 and \SI{1000}{Hz}).
By using only these \Omicron triggers, we can save the time that would otherwise be spent analysing data segments consistent with Gaussian noise; they do not teach us about the properties of glitches.
The inferred distribution of $\xiGH$ and $\xiGL$ given this contextual data is consistent with unity. This is not surprising since the \Omicron pipeline is designed to identify non-Gaussian noise.

For calculations of the astrophysical odds, we approximate the distribution of $\xiGH$ and $\xiGL$ using a point estimates $\xiGHhat$ and $\xiGLhat$ given by the ratio of the number of \Omicron triggers, for each detector, to the available data span. That is, we assume $\pi(\xi_{g} | d_{i\neq k}, I) = \delta\left(\xi_{g} - \hat{\xi}_{g}\right)$. The values of these point estimates are reported in Table~\ref{tab:overview}.
To verify that these point estimates are appropriate, we additionally analyse an auxiliary set of conditional data: 1000 randomly selected times near to GW151216. 
This contextual data has too few glitches to give reasonable inferences about $\lambda_\noise$, but gives a good measure of the glitch probability with medians and 90\% credible intervals $\xiGH=0.013^{+0.02}_{-0.01}$ and $\xiGL=0.0034^{+0.01}_{-0.003}$. 
The \Omicron rate estimates (Table~\ref{tab:overview}) lie at the 80\% and 96\% percentiles for the Hanford and Livingston detectors respectively. We conclude that the \Omicron triggers provide reliable point estimates, but that they are slightly conservative; by slightly overestimating $\xi_G$, there is a modest bias against the astrophysical hypothesis. In Sec.~\ref{sec:discusion} we show that the results are robust to this conservative choice.

When writing out the prior previously, each term was conditional on both the contextual data as well as $I$. However, by using the \Omicron triggers to infer $\lambda_\noise$, but point estimates to infer $\xiGH$ and $\xiGL$ we see that we are calculating
$\pi(\Lambda_\noise | d_{i\neq k}, I)=\pi(\xiGH | I) \pi (\xiGL | I)\pi\left(\lambda_\noise | d_{i\neq k}\right)$.

Having described details of our calculation, we now recap the procedure from start to finish.
There are three steps.
First, we identify a \SI{24}{\hour} period of data passing the standard data-quality vetoes and absent of injected signals and the analysis segment itself. Second, we filter the available data against \Omicron triggers to produce a list of contextual data segments known to contain glitches. Third, we analyse the loudest $N$ of these triggers and estimate the glitch hyper-parameters $\lambda_\noise$. In this step we vary $N$ by a factor of two and check that the resulting glitch population posteriors are invariant: this demonstrates that we have captured the typical glitch population properties without analysing the entire available data set. Finally, we calculate the astrophysical odds, Eq.~\eqref{eqn:bcr2}, using the distribution of hyper-parameters found in the second step, the prior odds $\langle \xi \rangle=7.4\times10^{-4}$, and the point estimates $\xiGHhat$ and $\xiGLhat$.

\section{Waveform models, priors, and noise uncertainty}
\label{sec:details}
We use the aligned-spin waveform model \texttt{IMRPhenomD} \citep{PhysRevD.93.044006, Khan:2015jqa} for the signal model and for the incoherent-between-detectors glitch model. 
In the future, it is desirable to extend this analysis to use more sophisticated waveforms including precession of the orbital plane and marginalization over systematic waveform uncertainties \citep{ashton2020}.
However, we elect to use \texttt{IMRPhenomD} because it is fast and no published candidate events exhibit strong evidence of precession.

We use data from the Gravitational Wave Open Science Centre \citep{gwosc} spanning $\unit[20-512]{Hz}$. We estimate the noise properties, the \ac{PSD}, from the median average of 31 non-overlapping \SI{4}s periodograms using \texttt{gwpy} \citep{2019ascl.soft12016M, duncan_macleod_2020_3598469}. The data used for estimating the \ac{PSD} is off-source and immediately before the analysis segment in each instance. We do not include the effects of calibration uncertainty \citep{Cahillane:2017jb}.

For signals, we use uniform priors in the chirp mass and mass ratio over the ranges $[13, 100$~$\Msun$ and $[0.125, 1]$ respectively; for the component spin prior we use the ``$z$-prior'' (see Eq. (A7) of \citet{lange2018}) which places much of the prior support at small spins; this is equivalent to the aligned-spin prior (Config.~B) used in \citet{Huang2020}. For the remaining parameters we use standard priors (see \citet{bilbyO1O2}), which are informed by the astrophysical nature of expected signals.
In the future, it is worth employing more realistic population models for mass and spin, though, this is outside our present scope; see~\cite{Fishbach,Galaudage2019}.

The informative prior distributions used for signals are not necessarily appropriate for the glitch model in which we project glitches into the compact binary coalescence signal parameter space.
The astrophysical odds framework is designed to use knowledge about typical glitches by marginalizing over the contextual data. 
It does so by ``recycling'' posteriors obtained with an initial prior (see Appendix B of \citet{bcr2}).
This process is inefficient if the glitch posteriors strongly disagree with the initial prior.
We find, in agreement with \citet{Davis2020}, that glitches tend to have posterior support in regions of parameter space unusual for typical astrophysical signals, e.g., large negative spins and extreme mass ratios. 
To counter this inefficiency, we apply, a glitch prior uniform in the component spin $\chi_1\in[-1, 1]$ and $\chi_2\in[-1, 1]$. 
In testing, we find this improved the efficiency of the astrophysical odds in properly classifying glitches. One might worry that, by applying a different prior for glitches and signals, we are biasing the odds. However, posterior samples are ultimately recycled using hierarchical inference, and so these prior choices do not affect our results except to improve computational efficiency.

We also find that the astrophysically motivated co-moving volumetric prior \citep{bilbyO1O2} for luminosity distance can also decrease the efficiency of recycling as most glitches tend to occur around $\sim 100$~Mpc. To be clear, glitches have no physical distance; we refer here to the effective distance obtained by fitting glitches to binary merger waveforms.
We therefore employ a uniform-in-luminosity distance prior for both signals and glitches, which ensures efficient recycling.

In testing, we find that it is important to include uncertainty in our estimate of the \ac{PSD} estimation.
Failing to take this into account yields false-positive signals ($\odds>1$) in time-slide checks in which the H1 data is offset from L1 to destroy the coherence of real gravitational-wave signals in the data.
The solution is to marginalise over uncertainty in the noise \ac{PSD} as in~\cite{Talbot2020, Banagiri2020}. Using the median Student-$t$ method from \cite{Talbot2020}), the astrophysical odds calculated for the set of time-slid \Omicron triggers behaves properly: all triggers result in an odds disfavouring an astrophysical interpretation (see Fig.~\ref{fig:plot}). 
We conclude that marginalizing over uncertainty in the \ac{PSD} is necessary for a reliable odds, and so we apply this to all the results discussed below.

\section{The glitch population}
\label{sec:glitches}
We infer the properties of the glitch population by analysing the top 100 \Omicron triggers from Hanford and Livingston in a \SI{24}{\hour} period around each of the events. Our glitch model consists of compact binary coalescence signals with uncorrelated parameters in each detector \citep{bcr2}.
As such, we are projecting the properties of glitches (for which we do not have a first-principle model) onto the parameter space of astrophysical signals. 
We find broadly consistent features in the glitch populations surrounding each of the three events.
Namely, glitches tend to have large anti-aligned spins $\chi \sim -1$ , large masses, and large mass ratios. 
These finding are consistent with the impulse-like glitches studied by \citet{Davis2020} for the \textsc{PyCBC} search algorithm~\citep{Nitz2017}.

For this analysis, we apply a simple glitch hyper-model to capture salient features of the triggers in our study. 
For each component spin, we apply a hyper-model consisting of a mixture distribution between a uniform and normal distribution. 
We verify the predictive power of the model by generating posterior predictive distributions and comparing these with the data. In future work, we will look to study a larger population of glitches and develop a more sophisticated glitch hyper model, potentially improving the ability of this method to distinguish astrophysical signals.

In this analysis, we study a limited set of contextual data to infer the properties of typical glitches. The conclusions are robust to this choice of contextual data and we find that doubling the number of points does not significantly change the inferred glitch population distribution. In future work, we will extend the analysis of glitches to a broader population and develop a more sophisticated glitch hyper-models. This will likely improve the ability of the method to distinguish astrophysical signals.

\begin{table*}
\renewcommand{\arraystretch}{1.2}
\setlength{\tabcolsep}{5pt}
\addtolength{\tabcolsep}{-1.5pt}
    \centering
    \caption{Significance estimates and associated quantities. The first five columns give $1-p_{\rm astro}$ for the GstLAL and PyCBC pipelines reported in \citet{GWTC1}, the PyCBC pipeline as reported in 1-OGC~\citep{1OGC} and 2-OGC~ \citep{nitz2020}, and results from the Institute for Advanced Study (IAS) \citep{Zackay2019, Venumadhav2019}. The next three columns give the prior probabilities for signals and glitches; see Sec.~\ref{sec:method}. The next five columns give the natural logarithm of: $B_{\rm S/N}^{G}$, the signal vs. Gaussian-noise Bayes factor; $B_{\rm S/N}$, the signal vs. Gaussian-noise Bayes factor marginalizing over uncertainty in the \ac{PSD} (see Sec.~\ref{sec:details}); $B_{\rm coh,inc}$, the coherent vs. incoherent Bayes factor \citep{veitch2010}); BCR, the Bayes coherence ratio \citep{isi2018}; and, in the second-to-last column, the astrophysical odds, $\odds$, (this work) marginalizing over prior probabilities (columns five to seven) and the $\chi_1,\chi_2$ glitch hyper-model. The final column is the remaining terrestrial probability where $\pastro \equiv \odds / (1 + \odds)$, this can be directly compared to the output of the search pipelines.}
    \begin{tabular}{l|ccccc|ccc|ccccc|c}
    \hline\hline
         Event & GstLAL & PyCBC & 1-OGC & 2-OGC & IAS & $\langle \xi \rangle$ & $\xiGHhat$ & $\xiGLhat$ & $\ln B^G_{\rm S/N}$ & $\ln B_{\rm S/N}$ & $\ln B_{\rm coh,inc}$ & $\ln$ BCR & $\ln \odds$ & $1 - \pastro$ \\ \hline
         GW150914 & $ <10^{-3} $ & $ <10^{-3} $  & $<8{\times}10^{-4}$ & $ <10^{-3} $ & -- & $7.4{\times}10^{-4}$ & 0.0094 & 0.013 & 307 & 205 & 12.5 & 14.3 & 16.2 & $9\times10^{-8}$\\
         GW151012 & $0.001$ & 0.04 & $0.024$ & $ <10^{-3} $ & -- & $7.4{\times}10^{-4}$ & 0.031 & 0.021 & 28.2 & 13.2 & 9.63 & 5.64 & 5.74 & 0.003\\
         GW151216 & -- & -- & 0.997 & 0.82 & 0.29 & $7.4{\times}10^{-4}$ & 0.022 & 0.016 & 12.7 & 3.70 & 3.10 & -3.53 & -3.50 & $0.97$\\
         \hline\hline
    \end{tabular}
    \label{tab:overview}
\end{table*}

\section{Results \& Discussion}
\label{sec:discusion}
We present the astrophysical odds for the three events analysed in this work in Table~\ref{tab:overview}. 
For GW151216, we find $\pastro=\result{}$ suggesting it is likely of terrestrial origin.
While the astrophysical odds disfavour an astrophysical origin, the posterior odds, $\mathcal{O}=\result{}$, are larger than the prior odds $7.4\times10^{-4}$, showing that this segment is 40 times more likely than average to contain a signal.
Our result suggests that the astrophysical implications inferred from GW151216 may be premised on a terrestrial event~\citep{piran2020,fragione2020,luca2020}.

In Table~\ref{tab:overview}, we also provide several Bayesian estimates of significance. These are the signal vs. Gaussian noise Bayes factor as measured directly $B_{\rm S/N}^{G}$, the signal vs. Gaussian noise Bayes factor after marginalizing over uncertainty in the \ac{PSD}; $B_{\rm S/N}$, the coherent vs. incoherent Bayes factor $B_{\rm coh,inc}$ \citep{veitch15}, and the Bayesian coherence ratio (BCR) \citep{isi2018}. The astrophysical odds builds on each of these concepts, as such, we can see each as a special case. $B_{\rm coh,inc}$, does not include the prior-odds and gives only the evidence for a signal vs. a glitch (i.e. the non-Gaussian component of the noise in our noise model). The BCR is an odds comparing the signal and noise hypotheses used in this work, but does not include the glitch hyper model marginalization.
In \citep{isi2018} the prior-odds and glitch probabilities are used to tune the statistic to maximise the detection power of the statistic in a bootstrap framework. In this work, we instead apply the usual direct interpretation for the tuning parameters as prior probabilities. In this sense, the astrophysical odds in the absence of a glitch hyper-model are equivalent to the BCR up to the choice of tuning parameters/prior probabilities. The difference between $\ln{\rm BCR}$ and $\ln \odds$ in Tab.~\ref{tab:overview} quantifies the effect of the glitch hyper model.

The astrophysical odds, as with any significance estimate, depends on the choice of priors and on the noise model.
It is therefore useful to consider which aspects of the analysis are most important for the conclusion that GW151216 is not astrophysical in origin.
From Table~\ref{tab:overview}, it is clear that GW151216 is a less significant trigger than the other two candidates from $B_{\rm S/N}^{G}$ alone; it has a lower signal-to-noise ratio.
However, the critical factor in our analysis responsible for reducing the significance of this event is the marginalization over the uncertainty in the \ac{PSD} (see Sec.~\ref{sec:details}).
In Table~\ref{tab:overview}, we see that the signal/noise Bayes factor falls from $B_{\rm S/N}^{G}=12.7$ for Gaussian noise to $B_{\rm S/N}=3.70$ using the Student-$t$ likelihood.

Naively combining this Bayes factor (ignoring the effect of glitches) with a prior odds of $\ln \langle \xi \rangle = -7.2$ the resulting odds is less than unity, providing evidence against an astrophysical origin. 
The subsequent BCR and astrophysical odds (which include the effect of this prior odds) make minor corrections, but retain the overall conclusion.
As discussed in Section~\ref{sec:details}, we cannot neglect this marginalization if we want reliable odds. This underlines the importance of \ac{PSD} estimation (for further discussion, see also \citet{Chatziioannou2019}).
In the future, with improved methods for evaluating the uncertainty in the \ac{PSD} (for example, building on the work of \citet{Biscoveanu2020} or developing a joint \ac{PSD} and model method building on \citet{2015PhRvD..91h4034L}), we can reassess GW151216. 

We now discuss the prior sensitivity of our results. The dominant prior choice is that of the $\xi$-distribution. In this work, we use an astrophysical prior based on the rate of binary black hole events in the O1 and O2 observing runs. The factorisation of the prior odds in Eq.~\eqref{eqn:bcr2} allows us to update the odds based on differing prior assumptions. In order to  change the conclusions for GW151216, one would need to increase $\langle \xi \rangle$ by a factor of $\sim 36$ Translating this into an updated merger rate, this would require a merger rate of $R'\sim 1600$~\si{\per\giga\parsec\cubed\per\year}, much larger than the current uncertainty on the merger rate \citep{GWTC1}. Similarly, a merger rate which would make GW151012 not of astrophysical origin (based on an updated prior odds) would also require a merger rate well outside of the current uncertainty. This demonstrates that our results are not sensitive to the choice of prior odds, given the current uncertainty. Technically, the odds for GW151012 and GW150914 are biased because the data from these events is used to estimate the rate. However, we expect the error from this double-counting to be negligible. The other potential bias from our prior assumptions is the choice of point estimates for $\xiGH$ and $\xiGL$. To check how sensitive our results are to this choice, we rerun the analysis of GW151216 using $\xiGH= \xiGL=0$ and find that $\ln \odds = -3.5$. This small shift from our calculated value confirms that our conclusion, that $\pastro=\result{}$, is robust to the choice of glitch hyper prior.

For GW150914 and GW151012, the astrophysical odds provide unequivocal evidence that these events are of astrophysical origin. Comparing the BCR and the $\odds$ in Table~\ref{tab:overview} allows to assess the effect of the glitch hyper-model.
For GW151012 and GW151216, only a small effect is observed, but for GW150914 the astrophysical odds is larger than the BCR by a factor of $\approx 7$. 
This demonstrates the ability of the astrophysical odds to increase our confidence in a signal based on how unlike the glitch population it is.
We can also compare the BCR values derived in this work with that of \citet{isi2018}. 
For GW150914 and GW151012, they find $\ln{\rm BCR}$ values of 19.6 and 8.5 respectively; larger than the values found in this work (see Tab.~\ref{tab:overview}). 
This difference is caused by an unknown combination of the choice of tuning parameters, the narrower source parameter priors, the use of a precessing waveform, or the marginalization over the \ac{PSD} applied in this work. 
Given the significant impact of the marginalization over the \ac{PSD}, we suspect this is likely to dominate, but we cannot determine this without further investigation.

\begin{figure}
    \centering
    \includegraphics{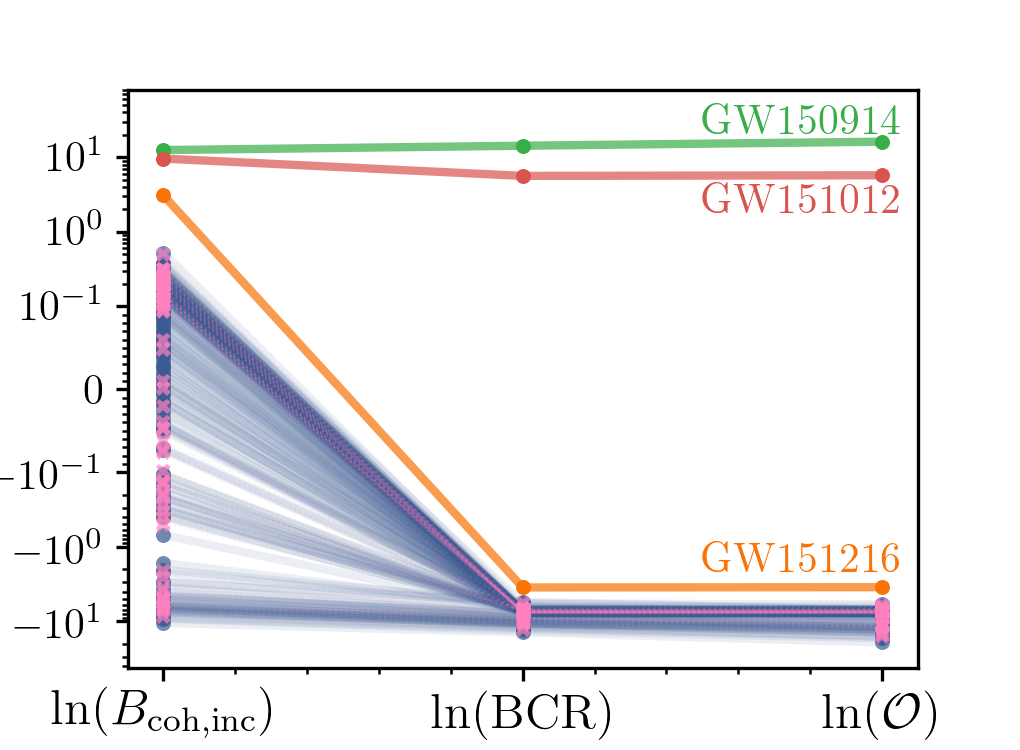}
    \caption{Visualisation of candidates and triggers considered in this work for three Bayesian significance estimates: $B_{\rm coh,inc}$, the coherent vs. incoherent Bayes factor \citep{veitch2010}; the ``Bayes Coherence Ratio'' (BCR) an odds comparing a signal with both incoherent glitches and Gaussian noise \citep{isi2018}; and finally the astrophysical odds, Eq.~\eqref{eqn:bcr2}. We label the three events analysed in this work in the figure and provide the numerical values for each in Tab.~\ref{tab:overview}. Blue circles and connecting curves are drawn for each of the \Omicron triggers used to characterise the background. Pink crosses and dashed connecting curves mark the values for time-shifted \Omicron trigger results---a background where we can be sure there are no coherent signals.}
    \label{fig:plot}
\end{figure}

To visualise our results for the three candidates and various realisations of a background, in Fig.~\ref{fig:plot}, we show the evolution of candidates through three stages of Bayesian significance estimates.
Individual candidates are labeled by their ID. 
In blue, are the \Omicron triggers identified for each of the three epochs around each event; we show these together as no differences in behaviour per-epoch were found. 
All the significance estimates use evidence obtained by marginalizing over the uncertainty in the \ac{PSD}.
For the \Omicron trigger candidates, we see two distinct clusters: those with $\ln(B_{\rm coh, inc})\sim 0$ and those with $\ln(B_{\rm coh, inc}) < -1$. These can be understood as a cluster of candidates where the data is reasonably Gaussian in both detectors (thus tricking the coherent Bayes factor which only compares signal evidence against glitch evidence) and a cluster of candidates with a strong glitch in one detector resulting in a Bayes factor favouring the glitch hypothesis. When subsequently analysed with the BCR metric \citep{isi2018}, the Gaussian cluster is weighted down because the BCR includes Gaussian noise in its alternative hypothesis. Finally, when applying the glitch hyper-prior a small correction is applied based on the likeness of the candidates to the glitch population. For the candidates initially in the $\ln(B_{\rm coh, inc}) < -1$ cluster, this results in a modest down-weighting: i.e. the odds having marginalized over the glitch population are slightly better at distinguishing glitches. In future work, we expect that a more detailed glitch model will yield further improvement in the ability of the odds to distinguish glitches.

In pink, we also show the evolution of a set of \Omicron triggers analysed with a time-slide. That is, we take the set of triggers and apply a \SI{1}{\second} shift between the Hanford and Livingston data. This ensures that the set of triggers do not contain coherent astrophysical signals. The figure demonstrates that the three Bayesian significance estimates perform equivalently for the \Omicron triggers under a time-slide as they do without. 

\section{Conclusion}
We find that the marginal gravitational wave candidate GW151216 is not of astrophysical origin, $\pastro=\result{}$. 
Our $p_\text{astro}$ estimate is smaller than that of the original detection claim $\pastro=0.71$ \citep{Zackay2019}, or the PyCBC analysis $\pastro=0.18$ \citep{nitz2020}. 
Taken together with \citep{Huang2020}, we urge the community to use caution when considering the astrophysical implications of this event.
We also analyse GW150914, the loudest signal in the first advanced-LIGO observing run, and GW151012, a candidate first marked as marginal, but subsequently upgraded. We find overwhelming support that these are astrophysical signals.

This work lays out the framework for applying the astrophysical odds \citep{bcr2} to a growing catalogue of gravitational-wave transients.
In doing so, we seek to provide a single $p_\text{astro}$ for candidate events from multiple groups.
Our results do not rely on the output of a search pipeline, and it is easy to see the assumptions that go into our calculations.
It is also straightforward to update our significance estimates to keep pace with advances in noise modelling.
Unlike traditional search methods, it does not use bootstrap realisations of the noise, but models the noise as incoherent-between-detector signals. 
In future work, we anticipate a number of improvements including: adding additional alternative models, for example, sine-Gaussians; improved waveforms; improved methods of estimating the noise \ac{PSD}; and the addition of calibration uncertainty. 

\section{Acknowledgements}
The authors are grateful to Jess McIver, Will Farr, Laura Nutall, Sebastian Khan, Max Isi, Thomas Massinger, Thomas Dent, our anonymous referee for useful comments during the development of this work.
The authors are grateful for computational resources provided by the LIGO Laboratory and supported by National Science Foundation Grants PHY-0757058 and PHY-0823459.
We acknowledge the support of the Australian Research Council through grants CE170100004, FT150100281, and DP180103155.
We use the \texttt{bilby} \citep{bilby} inference package and the \texttt{dynesty} \citep{dynesty} Nested Sampling algorithm.

\section{Data Availability}
This research has made use of data, and web tools obtained from the Gravitational Wave Open Science Center (GWOSC: \url{https://www.gw-openscience.org}), a service of LIGO Laboratory, the LIGO Scientific Collaboration and the Virgo Collaboration. LIGO is funded by the U.S. National Science Foundation. Virgo is funded by the French Centre National de Recherche Scientifique (CNRS), the Italian Istituto Nazionale della Fisica Nucleare (INFN) and the Dutch Nikhef, with contributions by Polish and Hungarian institutes. The data underlying this article are available from GWOSC (\cite{gwosc}) at \url{https://doi.org/10.7935/K57P8W9D}.

\bibliographystyle{mnras}
\bibliography{bibliography}

\bsp	
\label{lastpage}
\end{document}